Adoption of Social CRM in  Micro and Small Enterprises: An Analysis of Santarém's Market

Gustavo Nogueira de Sousa, 0000-0002-1517-3575, (Universidade Federal do Oeste do Pará, Pará, Brasil) - sougusta@gmail.com
Luan Vinícius Huppes, 0000-0003-3598-3708, (Universidade Federal do Oeste do Pará, Pará, Brasil) - luanviniciuspessoal@gmail.com
Antônio Fernando Lavareda Jacob Jr, 0000-0002-9415-7265, (Universidade Estadual do Maranhão, Maranhão, Brasil) - jacobjr@engcomp.uema.br
Fábio Manoel França Lobato, 0000-0002-6282-0368, (Universidade Federal do Oeste do Pará, Pará, Brasil) - fabio.lobato@ufopa.edu.br

**Abstract:** Online social networks have changed the ways of communication and social interactions, especially in the Customer Relationship Management (CRM). In this sense, a new concept about business strategies involving CRM and social media has aroused, known as Social Customer Relationship Management. Despite to be an emergent and promising research field, it was perceived that Micro and Small Enterprises (MSE) have shown few or no process of Social CRM implemented. Aiming to test this hypothesis, this work conducts a market analysis in Santarém City, located in the Pará State; evaluating the adoption of Social CRM by MSE. The main contribution of this study is related to the understanding of the dynamics between Social CRM and MSE. As results, the construction of insights' list of products and solutions suitable for the implementation of Social CRM by MSE, with the potential to guide research and development projects in this area.

Adoção de Social CRM em Micro e Pequenas Empresas: Uma Análise do Mercado Santareno

**Resumo:** As Redes Sociais Online (RSO) transformaram a forma de comunicação e de interação social, sobretudo no relacionamento entre empresa e consumidores. Neste sentido, surgiu um novo conceito acerca de estratégia de negócios envolvendo a gerência do relacionamento com o cliente e as mídias digitais, conhecido por *Social Customer Relationship Management* (Social CRM). Apesar de emergente e promissor, percebeu-se empiricamente que Micro e Pequenas Empresas (MPE) apresentavam poucos ou nenhum processo de Social CRM implementados. Visando testar esta hipótese, este trabalho realiza uma análise do mercado da cidade de Santarém - Pará, avaliando a adoção do Social CRM por MPE. A principal contribuição do estudo conduzido reside no entendimento da dinâmica entre Social CRM e o nicho empresarial em questão. O resultado foi a geração de *insights* sobre produtos e soluções adequados à implementação do Social CRM por MPE; com o potencial de guiar projetos de pesquisa e desenvolvimento na área.

**Palavras-chave:**  Social CRM, Micro e Pequenas Empresas, Redes Sociais Online, Sistemas de Informação, Estratégia de Negócio.

## 1. Introdução

O setor produtivo passou por mudanças drásticas com o surgimento das Redes Sociais Online (RSO), pois os consumidores têm se tornado cada vez mais engajados com este tipo de mídia colaborativa (Ahmad & Laroche, 2017). Isso implicou em uma grande mudança, onde as empresas estão redirecionando o seu foco para o relacionamento com o cliente, em que cada cliente passa a ter um tratamento de acordo com suas preferências e aspirações (Orenga-Roglá & Chalmeta, 2016).

Analisando as mídias sociais por meio do ponto de vista do consumidor, estas plataformas representam um canal para coleta de informação sobre as marcas e serviços, como também para emissão de opinião acerca deles (Constantinides & Holleschovsky, 2016). Já do ponto de vista das empresas, estas plataformas representam uma poderosa fonte de dados sobre o comportamento dos usuários, além de ser um canal de comunicação de baixo custo (Constantinides & Holleschovsky, 2016; de Almeida, Lobato, & Cirqueira, 2017; Lobato et al., 2017; Wang & Yu, 2015).

Observa-se ainda que esses usuários passaram de meros consumidores de informações, produtos e serviços, para *prosumers* (Ritzer & Jurgenson, 2010), consequentemente, estratégias tradicionais de gerenciamento de relacionamento com o cliente não funcionam de maneira efetiva (Lobato, Pinheiro, & Jr, 2017). Tal cenário fez surgir uma profusão de estratégias para se realizar a gerência de relacionamento com os clientes por meio das novas Tecnologias de Informação e Comunicação (TICs) (Soltani & Navimipour, 2016).

Este novo cenário impôs uma forte pressão mercadológica para as companhias aderissem à essa tendência, fazendo com que um outro conceito surgisse, o *Social Customer Relationship Management* (Social CRM), que visa a integração das redes sociais aos sistemas de CRM tradicionais (Alt & Reinhold, 2012; Kubina & Lendel, 2015). Entretanto, uma mudança na cultura corporativa passa a ser necessária para que a empresa enfrente os desafios do mercado e ganhe a vantagem competitiva por meio das relações sociais (Saba, 2013).

Micro e Pequenas Empresas (MPE), possuem a vantagem de estar em proximidade com os clientes o que cria um conhecimento claro de suas necessidades. Geralmente MPE não conta com hierarquias e burocracias bem definidas, o que significa flexibilidade para implementar mudanças (Harrigan & Miles, 2014). Na esfera econômica, de acordo com o (SEBRAE, 2015) em 2011 a participação de micro e pequenas empresas na economia brasileira foi de 27%. Sendo que no setor de serviço apresenta uma participação de 36,3%, setor de comércio de 53,4% e o setor industrial de 22,5%. Em momentos de crise político-econômica como a que se iniciou em 2014, percebe-se um aumento significativo da participação das pequenas e médias empresas na economia nacional.

Apesar de sua relevância econômica e relativa receptividade e flexibilidade para implementação de mudanças, como as requeridas pelo Social CRM, especialistas do Serviço Brasileiro de Apoio às Micro e Pequenas Empresas (SEBRAE) - Regional Baixo Amazonas, reportaram a baixa adoção de mídias digitais para gerenciamento de relacionamento com o cliente por este extrato empresarial. Este fato foi corroborado por uma pesquisa exploratória que avaliou o padrão de utilização de redes sociais por 120 empresas do setor do entretenimento, das quais apenas 54 possuíam perfis em redes sociais, mas apenas 49 empresas utilizavam tais plataformas de forma satisfatória e atendiam aos critérios de seleção - esta pesquisa será apresentada adiante.

Diante desta situação, nós hipotetizamos que restrições orçamentárias, de recursos humanos e falta de conhecimento são, muitas das vezes, uma barreira para a implementação de estratégias Social CRM por micro e pequenas empresas. Portanto, definiu-se como objetivo do trabalho a análise da dinâmica da adoção e implementação de Social CRM por micro e pequenas empresas. Sendo assim, as seguintes perguntas de pesquisa foram levantadas:

- PP1: Qual o perfil de utilização de redes sociais por micro e pequenas empresas?

- PP2: Quais as necessidades de produtos e serviços que micro e pequenas empresas apresentam para implementar o Social CRM?

A investigação para responder às perguntas de pesquisa acima descritas foi conduzida sob a ótica do *Design Science Research* (DSR) (Peffers, K., Tuunanen, T., Rothenberger, M. A., & Chatterjee, 2007), visando a produção de uma radiografia do cenário mercadológico do Social CRM para micro e pequenas empresas, mais especificamente, buscando entender a realidade do uso das redes sociais para o gerenciamento de clientes neste este nicho de mercado.

Para tal, selecionaram-se 49 empresas do setor do entretenimento da cidade de Santarém com representatividade de mercado. Este setor foi escolhido considerando que o Turismo é uma das principais atividades econômica da cidade. As empresas selecionadas foram analisadas quanto ao uso das redes sociais (e.g. presença nas redes, frequência e conteúdo de postagens, *etc*). Posteriormente, conduziu-se entrevistas estruturadas em 15 empresas a fim de avaliar quantitativamente determinadas variáveis, ao mesmo passo que atributos qualitativos também eram considerados.

Os resultados obtidos permitem responder às perguntas levantadas de forma satisfatória e confiável, como impacto imediato, destaca-se a construção de uma lista de necessidades e desafios que as micro e pequenas empresas possuem para implementar efetivamente os processos de Social CRM em ambiente de produção.

O restante deste artigo está estruturado como segue. A Seção 2 apresenta uma breve fundamentação teórica sobre micro e pequenas empresas e Social CRM; A Seção 3 descreve a metodologia adotada, seguido pelos resultados e discussões apresentados na Seção 4. Por fim, as conclusões e direcionamentos futuros são expostos na Seção 5.

**2. Fundamentação teórica**

Nesta seção a classe de micro e pequenas empresas é descrita, bem como a gerência de relacionamento com os clientes e sua variante baseada em redes sociais. A seção é finalizada com a apresentação e discussão dos principais desafios para a efetiva implementação do Social CRM em ambientes corporativos.

**2.1 Micro e pequenas Empresas**

No Art. 3º da Lei Complementar n.º 123 de 14 de dezembro de 2006 (Planalto, 2006) é considerado microempresas ou empresas de pequeno porte, a sociedade empresária, a sociedade simples, a empresa individual de responsabilidade limitada e o empresário, que estão devidamente registrados no Registro de Empresas Mercantis ou no Registro Civil de Pessoas Jurídicas e tenham receita bruta anual de igual ou inferior a R$ 360.000,00 (trezentos e sessenta mil reais); para que uma empresa seja considerada de pequeno porte a receita bruta anual deve ser superior a R$ 360.000,00 (trezentos e sessenta mil reais) e igual ou inferior a R$ 4.800.000,00 (quatro milhões e oitocentos mil reais).

O relatório desenvolvido pela (CNDL & SPC Brasil, 2015) nos revela algumas características quanto ao perfil do empreendedor que opera uma micro e pequena empresa. Cerca de 61% dos proprietários são homens, 54% têm idade entre 35 e 54 anos, 46,2% possuem renda familiar de 3 a 10 salários mínimos e aproximadamente 4 em cada dez gestores ou 39,8%, possuem ensino superior completo ou pós-graduação/especialização/MBA. Além disso, em média, um micro ou pequena empresa emprega seis funcionários.

O estudo conclui que, gestores com grau maior de instrução tem o faturamento cerca de sete vezes maior, reforçando a ideia de que a capacitação dos gestores é elemento fundamental para o sucesso do empreendimento. Os dados apresentados nesse relatório são de grande relevância para o planejamento estratégico de produtos ou serviços com foco em micro e pequenas empresas.

### 2.2. *Customer Relationship Management* (CRM)

O conceito de CRM é um termo da indústria da informação, usado para metodologias e software que combinam estratégias e tecnologias com o foco no cliente, que visa relacionamentos rentáveis a longo prazo e oferece meios para entender melhor e com detalhes suficientes os clientes, a fim de que a gerência, os vendedores e prestadores de serviços possam combinar as necessidades do cliente a planos e ofertas de produtos. Os sistemas de CRM estabelecem perfis de clientes de acordo com dados de ação, dados de reação, dados pessoais e dados potenciais de interações passadas (Wittwer, Reinhold, & Alt, 2017; Xu, Yen, Lin, & Chou, 2002).

O sistema de CRM é constituído de três subsistemas de acordo com (Tohidi & Jabbari, 2012b), a saber: i) o **CRM operacional**, que é o software que lida com o cliente face a face, como a automatização de venda, automação de marketing comercial, automação de suporte e proteção do cliente; ii) o **CRM analítico**, responsável por receber, armazenar, extrair, interpretar e gerar relatórios a partir dos dados do cliente; e iii) o **CRM relacional** no qual todas as formas de relação da organização com o consumidor com e-mails, telefone e entre outras são armazenadas e cruzadas.

Seguindo este raciocínio, podemos dizer que o CRM é um conjunto de tecnologias e processos usadas para a aquisição, análise e produção de conhecimento sobre o consumidor, que insere em toda e qualquer interação com consumidores (Bose, 2002). Toda a sua abordagem se baseia em reter e fidelizar o consumidor por meio de um bom relacionamento, uma vez que um dos fatores que determinam o sucesso ou o fracasso das empresas é a capacidade de compreender as necessidades e oferecer serviços com valor agregado aos clientes (King & Burgess, 2008).

Dessa forma, além de uma simples ferramenta de negócio, o CRM se configura como uma estratégia de negócio, melhorando processos internos de marketing, vendas, serviços e suporte aos clientes, que coopera diretamente para o sucesso de uma empresa (Ling & Yen, 2001; Xu et al., 2002). É possível também integrar o setor marketing a outros processos, a fim de atender e responder às pressões do mercado, mediante a flexibilidade em atender às mudanças de necessidades com eficiência e eficácia (Chen & Popovich, 2003). Diante o surgimento e expansão das mídias sociais, o relacionamento com os consumidores passou a encarar um novo desafio, o de gerenciar o relacionamento por meio dessas plataformas, surgindo um novo conceito, o Social CRM (Alt & Reinhold, 2012).

### 2.3. Redes sociais e CRM

O Social CRM consiste em integrar a análise de redes sociais ao gerenciamento tradicional de relacionamento com o cliente (Lobato et al., 2017). Assim com o CRM, o Social CRM é tido como uma estratégia de negócios apoiada por uma plataforma de tecnologia, regras de negócios, processos e características sociais, com o foco em gerar uma ligação com o cliente por meio de relações fortes e produção de valor em um ambiente transparente e confiável (Alt & Reinhold, 2012; Faase, Helms, & Spruit, 2011).

É importante notar que não existe um único produto chamado Social CRM, mas um conjunto de serviços existentes integrados em um ambiente de CRM, que agrega, às ferramentas de vendas, fontes adicionais de informações (Faase et al., 2011). Técnicas de monitoramento e análise de conteúdo social são usadas e seus resultados são integradas aos processos do Social CRM (Wittwer et al., 2017).

As plataformas sociais que podem ser integradas são as mais diversas, desde redes sociais *mainstream* como Twitter, Facebook e Instagram até plataformas de *electronic word-of-mouth* como o ReclameAqui e Ebit (de Almeida, Lobato, & Cirqueira, 2017). Independente da rede, a extração de conhecimento para suporte à decisão se faz impreterível, dentre as incontáveis análises podemos destacar a identificação de sentimentos para a mineração de opiniões sobre produtos e serviços (Cirqueira et al., 2017), a segmentação de mercado por meio de detecção de comunidade (Silva, Santana, Lobato, & Pinheiro, 2017), e o marketing direcionado. Em (Lobato et al., 2017) os autores identificaram um portfólio de serviços que emergiram com as redes sociais de acordo com as áreas relacionadas ao CRM, este portfólio encontra-se descrito na Tabela 1.

**Tabela 1:** Potenciais análises e serviços vislumbrados no contexto do Social CRM, adaptado de (Lobato et al., 2017).

| Setores Relacionados ao CRM | Portfolio de potenciais serviços em Social CRM |
|---|---|
| Vendas | Recomendação de produtos |
| | Predição de compra |
| | Identificador de *leads* (consumidor em potencial) |
| Marketing | Análise de mercado |
| | Campanhas em mídias sociais |
| | Gerência de marca adaptativa |
| | Avaliação do impacto de campanhas de publicidade |
| Serviços & Suporte | FAQs em mídias sociais |
| | Atribuição automática de postagens ao setor competente |

|  | Fóruns de suporte à comunidade |
|---|---|
| Inovação | Redes sociais empresariais |
|  | Identificação dos desejos e necessidades dos consumidores |
|  | Identificação de tendências |
| Colaboração | Identificação de influenciadores digitais |
|  | Recrutamento de colaboradores |
| Experiência do consumidor | Recrutamento de embaixadores de marca |
|  | Identificação e fomento de comunidades de influenciadores |

Além do portfólio de serviços acima mencionados, os autores identificam ainda onze desafios para a implantação efetiva do Social CRM em cenários reais, os quais foram classificados em três categorias, como apresentado na Tabela 2.

**Tabela 2:** Desafios para implementação Social CRM em cenários reais. Adaptador de (Lobato et al., 2017).

| Categorias | Desafios do Social CRM | Descrição |
|---|---|---|
| Processos | Integrar CRM com mídias sociais | Muitas empresas não sabem como integrar o CRM com a mídia sociais. |
|  | Métricas | Indicadores-chave de performance específicos para o Social CRM devem ser usados para uma avaliação fidedigna. |
| Infraestrutura | Excelência operacional | Há a necessidade de capacitar os colaboradores no uso de ferramentas e estratégias de Social CRM, para que haja uma imersão total na cultura das mídias sociais. |
|  | Investimentos | A falta de recursos para pequenas e médias empresas é latente. Pelos desafios inerentes à mensuração do Retorno do Investimento (ROI) em mídias sociais, as pequenas e médias empresas tendem a ver tal campo como um custo e não como investimento. |

|  | Ferramentas | Os sistemas atuais são limitados à um determinado tipo de conteúdo de redes sociais e negligenciam grandes quantidades de dados oriundo de outras fontes. |
|---|---|---|
| Data Mining | Big Data | A análise de grande quantidade de dados é desafiadora, por suas características inerentes, como volume, velocidade e variedade de conteúdo (por exemplo, imagens, vídeo e texto em diferentes idiomas, às vezes com expressões particulares de cada idioma). |
|  | Tipos de dados | A heterogeneidade do tipo de dados em redes sociais representa um grande desafio para Social CRM, bem como conteúdo multimídia para técnicas de extração de conhecimento, que tem o custo computacional elevado. |
|  | Estruturas de dados | Aproximadamente 90% dos dados da rede social são não-estruturados, o que é um desafio para os métodos de mineração de dados tradicionais, geralmente aplicado a dados estruturados. No entanto, o processamento de dados não estruturados pode melhorar significativamente o Social CRM. |
|  | Segurança dos dados | Mecanismos de rastreamento e gerenciamento da confiabilidade, autenticidade e segurança do conteúdo gerado pelo usuário podem melhorar o desempenho das técnicas de Social CRM, mas isto é frequentemente negligenciado. |
|  | Privacidade | É necessário ética para coletar, processar, usar e reportar dados extraídos das redes sociais para que não haja violação da privacidade do cliente |
|  | Velocidade | Há uma demanda notória por sistemas e processos que permitam o monitoramento e análise de dados em tempo real. |

Apesar de todas as dificuldades encontradas para a implementação do Social CRM, micro e pequenas empresas podem usar facilmente a tecnologia de redes sociais para gerenciar o relacionamento com os clientes sociais. Sendo muito compatível com a estrutura comercial existente, uma vez que a tecnologia das redes sociais é muito simples e pode ser adotada por qualquer organização (Ahani, Rahim, & Nilashi, 2017). No entanto, a realidade brasileira impõe severas restrições de recursos, tanto humanas (e.g. pessoal capacitado), quanto orçamentárias, uma vez que a maior parte das ferramentas de monitoramento de redes sociais e de CRM são plataformas estrangeiras que cobram licença de uso em dólar. Este trabalho busca justamente identificar quais são os desafios a fim de destacar oportunidades de negócios na área.

## 3. Metodologia

O *Design Science Research* é uma metodologia de pesquisa bastante difundida quando o objetivo final do projeto é o desenvolvimento de um produto, metodologias, estratégias e serviços, sobretudo na área de Sistemas da Informação. Durante a construção deste projeto, percebeu-se que esta metodologia se adequa precisamente ao escopo de pesquisa deste trabalho. Composto de seis etapas postas em sequências, o DSR é um modelo de processo. No presente estudo adotaremos as definições e funções de cada sequência de acordo com (Peffers, K., Tuunanen, T., Rothenberger, M. A., & Chatterjee, 2007), sendo que a Figura 1 apresenta a instanciação das etapas do DSR para o projeto descrito no presente artigo.

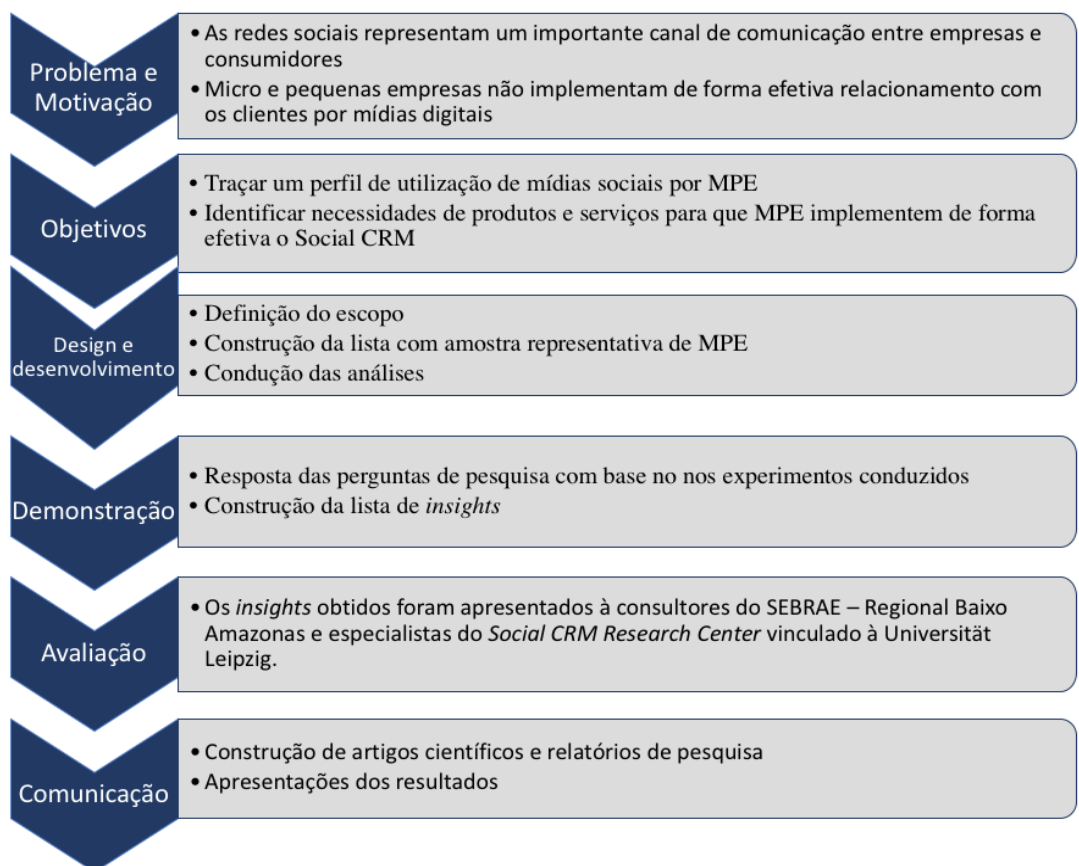

**Figura 1.** Instanciação das etapas DSR para o estudo em questão.

O primeiro passo apresentado na Figura 1 é a identificação do problema e motivação. Neste estudo, a identificação do problema adveio da percepção de mercado dos pesquisadores vinculados ao *Social CRM Research Center* da Universidade de Leipzig, Alemanha, durante a participação em feiras, *workshops* e conferências com pesquisadores e profissionais atuantes na área; e de reuniões com especialistas do SEBRAE Regional Baixo Amazonas.

Em um breve resumo, os consultores e pesquisadores atuantes em MPE afirmavam que a maior parte das ferramentas de CRM, análise de redes sociais e integradoras dos dados entre sistemas de CRM e redes sociais, são destinados a empresas médias e grandes - tanto no que tange à funcionalidade, mas sobretudo ao investimento para utilizá-las. Desse modo, as MPE sentiam-se excluídas desse mercado.

Outro ponto notório era em relação ao desconhecimento de práticas e processos de CRM (tradicional) e do Social CRM por parte dos gestores. Este fato, aliado às características da MPE apresentadas na seção anterior, foram motivadores deste estudo a fim de analisar este mercado com o intuito de se avaliar o grau de conhecimento/desconhecimento dos gestores. Desse modo, os objetivos de pesquisa foram delineados, tal como apresentado no segundo passo da Figura 1.

Acerca do *Design* e Desenvolvimento, esta etapa foi dividida em três fases. A primeira consistiu na definição do escopo de pesquisa, que se resume em micro e pequenas empresas ligadas ao setor do entretenimento. A motivação dessa escolha reside, sobretudo, na vocação turística da cidade onde a pesquisa foi conduzida, Santarém no Pará, sendo que este setor colabora significativamente para a economia regional. Ademais, nicho de mercado se beneficia diretamente das mídias sociais na forma *Business to Customer*, em detrimento de outros setores como agricultura e pecuária onde os negócios se baseiam em *Business to Business*.

Com a definição do escopo, construiu-se a lista de empresas-alvo e estabeleceram-se os critérios de exclusão, a saber:
- Não possuírem perfis em redes sociais;
- Empresas cujo perfis nas redes sociais não possuíam atividade nos últimos 12 meses;
- Estabelecimentos que não atendiam ao perfil de micro e pequenas empresas;
- Estabelecimentos cuja administração não era local, como por exemplo, rede de cinema;
- Empresas que fecharam entre a construção da lista e a análise das redes.

Após isso, foram realizados o planejamento e a condução das análises, as quais foram divididas em duas, uma exploratória, visando responder a PP1 e a outra por meio de um questionário aplicado à amostra das empresas a fim de responder a PP2. De posse das respostas, partiu-se para a quarta etapa da pesquisa, de demonstração, onde analisaram-se as respostas e geraram-se *insights* de oportunidades de P&D na área. Como os *insights* foram considerados significativamente relevantes, passou-se para a etapa de validação.

Neste ponto, os resultados e *insights* obtidos foram apresentados à consultores do SEBRAE - Regional Baixo Amazonas e especialistas em Social CRM do *Social CRM Research Center*, vinculado à Universidade de Leipzig. Todos validaram os achados e referendaram a metodologia adotada. Passada esta etapa, com os resultados validados, partiu-se para a última fase do DSR, de apresentação dos resultados, a qual culminou na construção deste artigo, cujos resultados das análises encontra-se descritos a seguir.

**4. Resultados e Discussões**

Nesta seção os resultados das análises são apresentados e discutidos. Ao final das análises mais relevantes, os achados são sumarizados.

**4.1 Definição do escopo e seleção da amostra**

O primeiro resultado da pesquisa foi a construção de uma lista contendo 120 empresas que se destacavam no setor em questão, incluindo bares, restaurantes, barbearias, cinemas, hotéis e casas de show. Desta primeira lista, alguns estabelecimentos foram removidos considerando critérios de exclusão expostos anteriormente. Com essa filtragem, apenas 49 empresas atendiam aos critérios de seleção.

**4.2 Perfil de utilização de mídias digitais**

**4.2.1 Presença nas redes**

Baseando-se nesta lista, a primeira análise foi conduzida visando responder a PP1: "Qual o perfil de utilização de redes sociais por micro e pequenas empresas?". Para tal, algumas estatísticas básicas foram computadas, como presença nas redes sociais mais utilizadas para Social CRM (Facebook, Twitter, Instagram e Tripadvisor); frequência de postagem; número de seguidores, *reviews,* dentre outros. Tais análises são apresentadas nas subseções a seguir. A Figura 2 apresenta um resumo sobre a presença em mídias digitais das empresas analisadas em números absolutos.

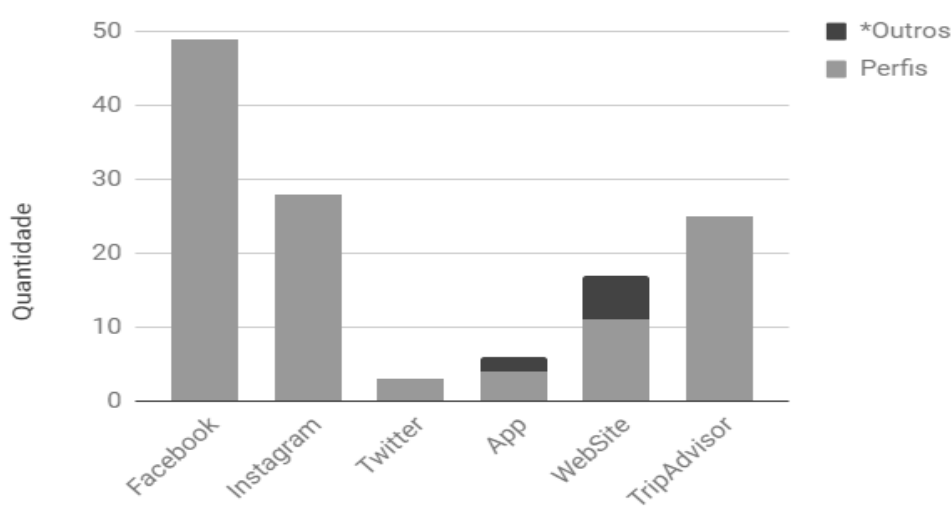

**Figura 2.** Quantitativo de perfis nas redes sociais de acordo com os estabelecimentos analisados.

Analisando a Figura 2 é possível perceber que todas as empresas analisadas possuíam ao menos um perfil nas redes, com o Facebook sendo a plataforma mais pervasiva, seguido do Instagram. Isso se deve aos critérios de inclusão adotados. Em App mensuramos a quantidade de estabelecimentos que possuíam aplicativo, onde "*Outros" significa a presença em aplicativos que não sejam proprietários, como iFood.

O resultado para esta variável mostrou que apenas quatro das empresas possuíam aplicativo próprio e duas empresas usavam o iFood, esta observação é discrepante com o senso comum em capitais, onde a presença em aplicativos terceirizados é mandatória (CNDL & SPC Brasil, 2017). Em uma análise mais aprofundada sobre o tema, percebeu-se que este é um aspecto cultural importante da cidade - a não aderência aos aplicativos.

Ainda acerca dos resultados dispostos na Figura 2, dezessete empresas possuíam website, sendo onze funcionais (perfil) e seis não funcionais (*Outros) - onde a empresa possuía o domínio, mas não o usava.

### 4.2.2 Padrão de Utilização

Considerando a presença a proporção de utilização do Facebook, Instagram e Tripadvisor, algumas estatísticas básicas foram calculadas para análise, conforme apresentado nas Tabelas 3, 4 e 5.

**Tabela 3.** Estatísticas de uso para o Facebook.

|  | Média de publicação por semana | Nº de curtidas | Nº de seguidores | Nota *Reviews* | Qtde *Reviews* |
|---|---|---|---|---|---|
| Média | 4,04 | 7.231,51 | 7.248,37 | 4,58 | 275,05 |
| Máximo | 24,38 | 35.301,00 | 35.296,00 | 5,00 | 2.005,00 |
| Mínimo | 0,00 | 625,00 | 648,00 | 3,80 | 25,00 |
| Desvio Padrão | 5,97 | 7.584,09 | 7.584,61 | 0,29 | 396,80 |

A Tabela 3 mostra as estatísticas resumo da utilização do Facebook para as 49 empresas analisadas. A média de publicação foi calculada considerando oito semanas de observação, sendo que a linha média mostra a média das médias de publicação. As outras variáveis analisadas são o número de curtida das páginas, número de seguidores, nota dos *reviews* e a quantidade dos *reviews*.

Alguns pontos chamaram a atenção dos pesquisadores durante a análise destes dados. O primeiro é que seis estabelecimentos não possuíam nenhuma postagem nas oito semanas analisadas, sendo que 34 estavam com menos de 4 postagens por semana. Outro fato interessante foi relacionado ao outro extremo, uma grande atividade na rede - havia um restaurante com a média de publicação por semana igual à 24, perfazendo cerca de 3,5 publicações por dia. No entanto, esta atividade frequente não implicou em um número maior de curtidas e seguidores (5.440 e 5.456, respectivamente), pelo contrário, tais variáveis encontravam-se abaixo da média; o mesmo é válido para o número de revisões - apenas 34.

Devido às características da plataforma, o número de seguidores acompanha o número de curtidas, sendo que os três estabelecimentos com maior número de curtidas pertenciam ao segmento de casas noturnas, com valores iguais à 21.474, 32.418 e 35.301. Em relação aos *reviews*, a menor nota foi atribuída à uma pizzaria, onde as principais reclamações eram relacionadas à demora na entrega do pedido.

Três fatos chamaram atenção na análise dos *reviews*, o primeiro é que a empresa com pior avaliação não respondia os *reviews*. Outro ponto é que as reclamações sobre atrasos na entrega estão bem distribuídas na linha temporal, portanto permitiu-nos concluir que o estabelecimento não leva em consideração os dados dispostos no Facebook para melhorar os serviços prestados. Por fim, seis estabelecimentos desabilitaram a opção de *reviews*.

As análises do Instagram foram basicamente em relação a frequência de postagem e o número de seguidores, conforme apresentado na Tabela 4.

**Tabela 4.** Estatísticas de uso para o Instagram.

|  | Média de publicação (8 semanas) | Nº Seguidores |
|---|---|---|
| Média | 4,04 | 6.642,22 |
| Máximo | 21,63 | 23.200,00 |
| Mínimo | 0,00 | 332,00 |
| Desvio Padrão | 4,85 | 6.454,19 |

Os estabelecimentos mais seguidos no Instagram são sorveterias e docerias, contrastando com as casas de show que representavam os mais seguidos no Facebook. Quanto à frequência de postagens, ela é a mesma para Facebook e Instagram, uma vez que a maior parte dos gerenciadores de mídias sociais duplicam as mesmas, apesar de que o público-alvo delas seja diferente. Em uma análise preliminar, percebeu-se que o gênero feminino é mais presente no Instagram, o que pode justificar que os estabelecimentos com maior número de seguidores sejam sorveterias e docerias. Um levantamento similar foi feito para o TripAdvisor, conforme apresentado na Tabela 5.

**Tabela 5.** Estatísticas de uso para o TripAdvisor.

|  | Nota *Reviews* | Quantidade *Reviews* |
|---|---|---|
| Média | 3,84 | 103,16 |
| Máximo | 5,00 | 500,00 |
| Mínimo | 0,00 | 0,00 |
| Desvio Padrão | 1,50 | 163,20 |

Como o TripAdvisor é uma plataforma ela é colaborativa com pouca ou nenhuma interação por parte dos estabelecimentos, alguns achados curiosos foram encontrados. Por exemplo, o estabelecimento com maior número de *reviews* é um restaurante cuja a última postagem no Facebook foi a cinco meses anteriores à data da coleta dos dados. No entanto, por ser um local turístico os usuários tendem a avaliá-lo.

### 4.2.3 Achados das análises

Os principais achados das análises foram:
- Grande número de postagens semanais no Facebook não quer dizer aumento no número de seguidores;
- Apesar do número de seguidores ser grande no Facebook, para alguns estabelecimentos a quantidade de *reviews* pode ser pequena;
- A quantidade de avaliação no Facebook mostra-se não ter relação com a taxa média de postagens;
- O número de seguidores no Facebook é influenciado de acordo com o setor no qual o negócio está inserido;
- Alguns setores se destacam mais no Facebook como as casas de show e outros mais no Instagram como as docerias;
- Mesmo sem presença ativa no Facebook e Instagram, empresas podem ter grandes quantidades de avaliações no TripAdvisor;

- As notas das avaliações do Tripadvisor estão de acordo com as apresentadas no Facebook.

Estes achados representam um resumo da resposta para a PP1 - "Qual o perfil de utilização de redes sociais por micro e pequenas empresas?".

**4.3 Gerenciamento das mídias sociais e do relacionamento com os clientes**

A segunda parte do estudo foi conduzido visando responder a PP2: " Quais as necessidades de produtos e serviços que micro e pequenas empresas apresentam para implementar o Social CRM?". Para tal, desenvolveu-se um questionário que foi divulgado para as empresas listadas nas fases anteriores. No entanto, mesmo com a divulgação e contato por telefone, apenas 3 empresas responderam. A fim de aumentar a amostra, modificou-se a estratégia adotada passando-se a marcar visitas ao estabelecimento, aproveitando o ensejo para aplicar o questionário. Tal mecanismo se mostrou efetivo pois 15 empresas responderam o instrumento de pesquisa.

**4.3.1 Características do Mercado**

Algumas variáveis de pesquisa serão suprimidas pois eles eram para avaliar os resultados da pesquisa exploratória. Todas as perguntas neste sentido confirmaram os achados descritos anteriormente. As demais variáveis de pesquisa são descritas a seguir.

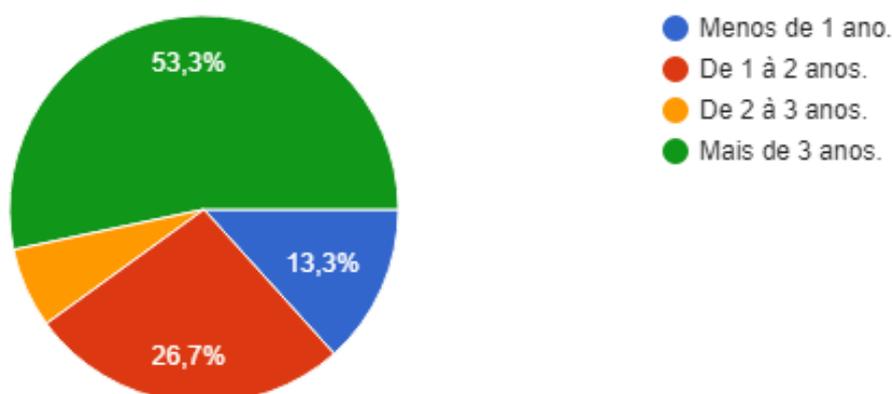

**Figura 3**. Tempo de mercado das empresas entrevistadas.

Como pode ser observado na Figura 3, 53,3% da amostra pesquisada possuem mais de 3 anos, no qual essa parcela de empresas já conta com uma certa estrutura e experiência de mercado. Em contrapartida, pode ser observado que cerca de 40% tem menos de 2 anos. Esse dado é muito útil para a aplicação de estratégias de Social CRM, visto que nos dá uma noção do nível de experiências das empresas presentes no mercado.

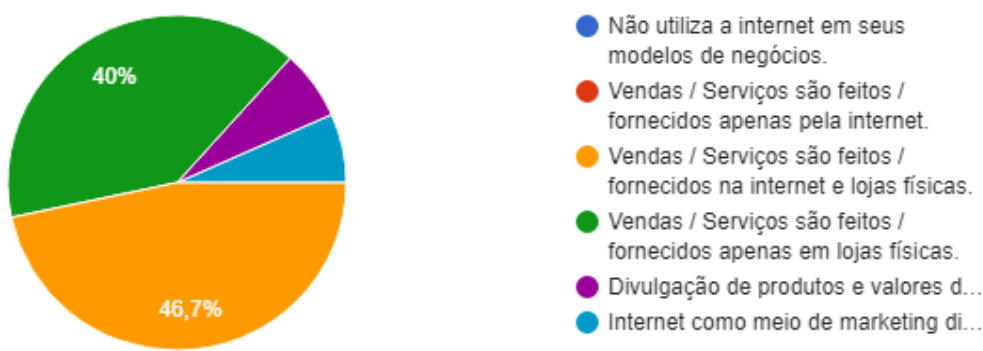

**Figura 4.** Modelo de negócios adotados pela empresa.

Podemos observar na figura 4 as diferentes formas de atuação no mercado pelas empresas, no qual 40% utilizam a internet para vendas de produtos ou serviço que são oferecidos apenas na física. Em torno de 46,7% utilizam a internet para vender produtos que são fornecidos pela internet e pela loja física. Podemos ver um certo equilíbrio entre esses dois tópicos, devido as necessidades mercadológicas de cada setor de entretenimento.

**4.3.2 Gerência das Redes Sociais**

Na Figura 5 é exposto a forma como se dá a gerência da RSO, no qual em 93,3% dos casos a gerência é feito pelo proprietário do estabelecimento. Esse fenômeno ocorre por ter somente micro e pequenas empresa na amostra, sendo que essas empresas não dispõem de recurso financeiros para dedicar um funcionário somente para o manejo das RSO.

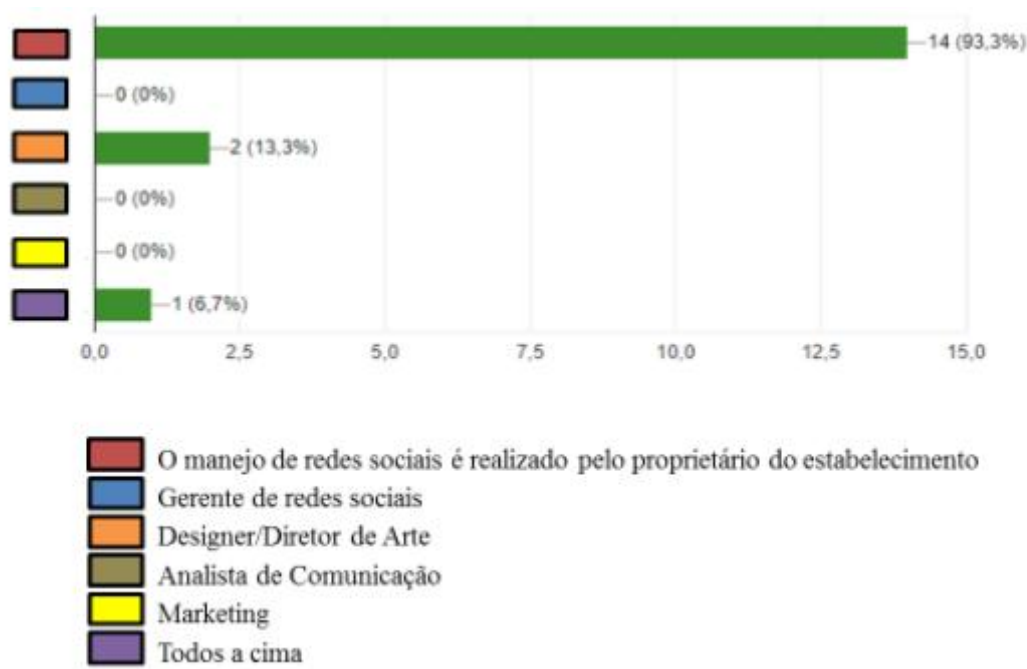

**Figura 5.** Cargos relacionados à rede sociais.

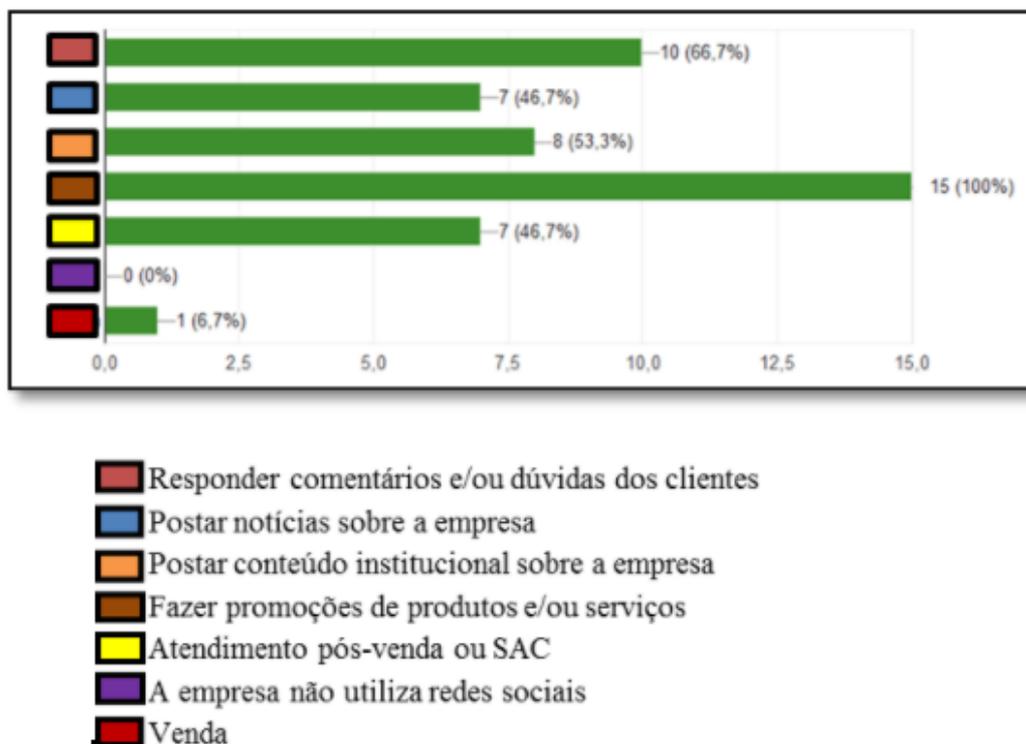

**Figura 6.** Objetivos de uso das RSO.

É possível observar na Figura 6 dados que indicam que 100% das empresas se utilizam das RSO para fazer promoções de produtos e/ou serviços, mas apenas 6,7% usam as RSOs como canal para vendas de produtos. Além disso vemos que 66,7% das empresas utilizam das RSO para responder comentários e/ou dúvidas dos clientes, o que é muito importante, pois como foi relatado anteriormente, empresas que ignoram seus clientes são mal avaliadas.

**4.3.3 Uso de CRM e adoção do Social CRM**

Por meio da análise da Figura 7 podemos observar certa deficiência no uso de sistemas de CRM, em que 80% das empresas não utilizam quaisquer tipos de sistemas de CRM, o que pode ocasionar prejuízos quanto a retenção do cliente. No entanto, um dado a ser destacado é o fato de uma das empresas utilizar o site próprio da empresa para pôr em prática atividades de CRM.

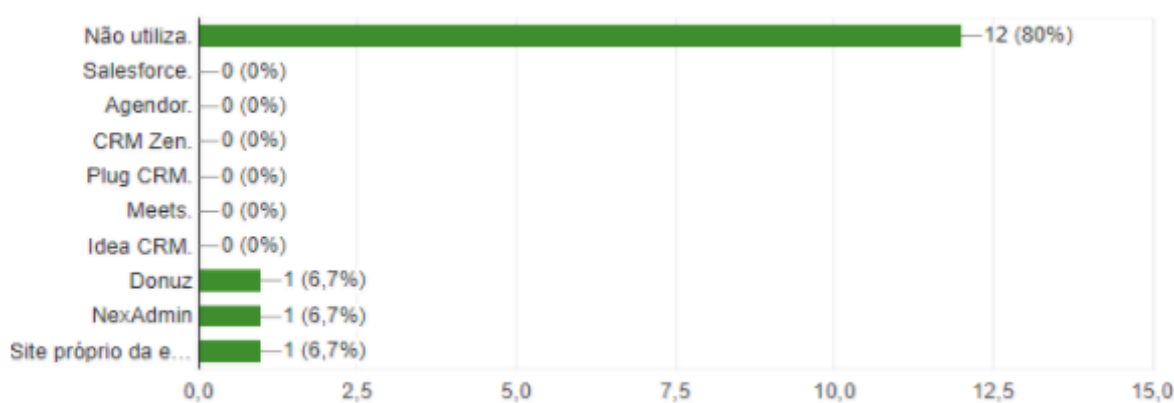

**Figura 7.** Uso de sistemas de CRM.

### 4.3.4 Achados da análise

- Apesar de terem um bom tempo de mercado, algumas empresas ainda têm grande deficiência no uso das RSO;
- A maior parte das MPE não tem um cargo específico para a gerência das RSO;
- O potencial das RSO não é completamente utilizado para o Social CRM pelas MPE;
- Divulgação de promoções para dias de pouco movimento é o principal atrativo das RSO para as empresas analisadas;
- Na prática, a maior parte das empresas consultadas não usam nenhum sistema de CRM;
- Em uma análise paralela, percebeu-se até a carência de sistemas de *Enterprise Resource Planning* (ERP).

### 4.4 Oportunidades de Pesquisa & Desenvolvimento

Considerando as análises conduzidas, esta seção descreve alguns *insights* quanto a oportunidades de pesquisa e desenvolvimento obtidos, respondendo assim à PP2. Primeiramente, por meio da análise exploratória das RSO percebeu-se a não adoção de boas práticas de Social CRM, perpassando pela má utilização de tais plataformas até o negligenciamento das redes sociais como mecanismos de comunicação/satisfação do cliente, como no caso das respostas aos *reviews*.

Em relação ao levantamento realizado na segunda análise, percebe-se que mesmo com relativo tempo de mercado, acima dos três anos, as empresas utilizam as mídias digitais apenas para divulgar produtos e serviços, não penetrando no mercado eletrônico. Ainda neste tocante, o pós-venda é pouco explorado.

Como o instrumento de pesquisa utilizado foi aplicado presencialmente na maior parte dos casos, uma das hipóteses levantadas por meio da observação das respostas é que há um desconhecimento de práticas e estratégias para se gerenciar o relacionamento com os clientes por meio das RSO. Sendo assim, a principal oportunidade vislumbrada é o fornecimento de oficinas e cursos de capacitação aos gestores a fim de informar as potencialidades do Social CRM. Outro ponto notório foi a da pouca utilização de sistemas de CRM, este quesito abre o precedente para oferecer serviços de implantação e testes de tais plataformas em ambiente de produção. Além disso, reforça-se a necessidade capacitação na área.

## 5. Considerações finais

É inegável o impacto das redes sociais online nas relações sociais, inclusive entre empresas e consumidores. Uma das consequências mais notórias aos negócios foi o surgimento de novas estratégias envolvendo a gerência de relacionamento com os clientes e as mídias digitais, conhecido por *Social CRM.*

Apesar das potencialidades da área, percebe-se que Micro e Pequenas Empresas estão à margem desta tendência. Visando melhor analisar este fato, uma análise de mercado ampla foi conduzida no presente estudo, a qual dividiu-se em duas etapas, uma exploratória com o objetivo de identificar o perfil de utilização de redes sociais por micro e pequenas empresas; e outra quantitativa, visando identificar e entender as necessidades de produtos e serviços que micro e pequenas empresas para implementar o Social CRM.

Os resultados obtidos permitem-nos concluir que apesar das empresas estarem presentes nas RSO por meio de seus perfis, a adoção de estratégias de Social CRM é extremamente baixa. Além disso pode-se perceber que a falta de conhecimento e as restrições financeiras estão entre os principais fatores para que isso ocorra. Esta pesquisa fornece dados que a tem impacto direto nas MPE, pois conhecendo e caracterizando a realidade de cada MPE, surge a possibilidade de fornecer capacitação em Social CRM para os gestores, bem como serviços de implantação e teste de plataformas de CRM.

Os desdobramentos desta pesquisa são variados. Em um primeiro momento pensa-se na expansão do número de empresas avaliadas e análise por setor econômico. Em um segundo momento considera-se relevante dedicar-se a investigações qualitativas, como por exemplo, analisar o padrão das postagens nas redes sociais, categorizando-as em divulgação, promoções e sorteios, por exemplo - a partir deste ponto será possível ver quais categorias possuem maior engajamento por parte dos usuários; e conduzir estudos de caso para entender melhor as necessidades das micro e pequenas empresas no tocante à implementação do Social CRM.